\documentclass[letterpaper, 10 pt, conference]{ieeeconf}  

\IEEEoverridecommandlockouts                              

\usepackage{amsmath}
\usepackage{amssymb}
\usepackage{mathtools}
\usepackage{nicefrac}
\usepackage{algorithm}
\usepackage{algpseudocode}
\usepackage[T1]{fontenc}
\usepackage{bm}
\usepackage{cleveref}
\usepackage{siunitx}
\usepackage{xcolor}
\usepackage{pgfplots}
\usepackage{pgfplotstable}
\usepackage{tikz-3dplot}
\usepackage{tikz}
\usepackage{expl3}
\usepackage{caption,subcaption}  
\pgfplotsset{compat=1.18}
\usepgfplotslibrary{groupplots}
\usetikzlibrary{decorations.markings}
\usetikzlibrary{external,calc}
\usepackage{xspace}
\usepackage[all]{nowidow}

\usepackage{amsthm}
\theoremstyle{remark}
\newtheorem{remark}{Remark} 

\definecolor{horizon_orange}{RGB}{252, 109, 35}
\definecolor{horizon_blue}{RGB}{120, 120, 255}
\definecolor{itm_blue}{RGB}{58, 152, 242} 

\definecolor{myblue}{RGB}{58, 152, 242} 
\definecolor{myorange}{RGB}{242,153,74}
\definecolor{mygreen}{RGB}{39,174,96}
\definecolor{mypurple}{RGB}{108, 92, 231}
\definecolor{mydarkgrey}{RGB}{52,73,94}

\tikzstyle{Rectangle}=[draw=black, shape=rectangle, minimum height=1.25cm, minimum width=2cm, align=center]
\tikzstyle{Circle}=[draw=black, shape=circle]

\tikzstyle{Arrow}=[->, >={Latex[scale=1.25]}]
\tikzstyle{Dashed Line}=[-, dashed, thick, black]

\newcommand{\drawcrazyflie}[4]{%

    \def\bladeLength{0.06}
    \def\bladeWidth{0.015}

    \pgfmathsetmacro{\xpos}{#1}
    \pgfmathsetmacro{\ypos}{#2}
    \pgfmathsetmacro{\angle}{#3}

    \begingroup\edef\temp{%
      \endgroup\noexpand\node[anchor=center, inner sep=0] at (axis cs:\xpos,\ypos) {\includegraphics[scale=#4, angle=#3]{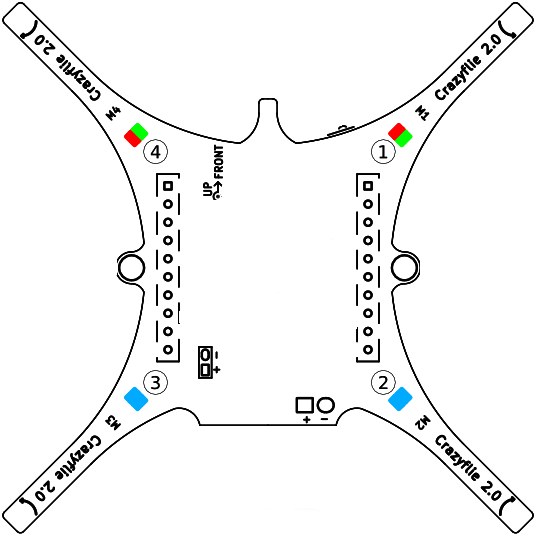}};
    }\temp

    \foreach \bx/\by/\ang in {
        0.04/-0.04/45,
        -0.04/-0.04/-45,
        -0.04/0.04/45,
        0.04/0.04/-45
    }{
        \pgfmathsetmacro{\xpos}{#1 + \bx*cos(\angle) - \by*sin(\angle)}
        \pgfmathsetmacro{\ypos}{#2 + \by*cos(\angle) + \bx*sin(\angle)}
        \pgfmathsetmacro{\angle}{\angle + \ang}
        \begingroup\edef\temp{%
          \endgroup\noexpand\filldraw[black] (axis cs:\xpos,\ypos) circle (0.0025);
        }\temp
        \begingroup\edef\temp{
          \endgroup\noexpand\draw[rotate around={\angle:(axis cs:\xpos,\ypos)}, fill=black]
              (axis cs:\xpos,\ypos)
                .. controls (axis cs:{\xpos - \bladeLength/2},{\ypos + \bladeWidth/2})
                          and (axis cs:{\xpos - \bladeLength/2},{\ypos - \bladeWidth/2})
                  .. (axis cs:\xpos,\ypos)
                .. controls (axis cs:{\xpos + \bladeLength/2},{\ypos - \bladeWidth/2})
                          and (axis cs:{\xpos + \bladeLength/2},{\ypos + \bladeWidth/2})
                  .. (axis cs:\xpos,\ypos);
        }\temp    
    }
}

\newcommand{\tran}{^{\mkern-1.5mu\mathsf{T}}}

\newcommand{\Crazyflie}{\textit{Crazyflie 2.1}\xspace}
\newcommand{\CrazyflieShort}{\textit{Crazyflie}\xspace}

\title{\LARGE \bf Efficient Collision-Avoidance Constraints for Ellipsoidal Obstacles in Optimal Control: Application to Path-Following MPC and UAVs
}

\author{David Leprich$^{1}$, Mario Rosenfelder$^{1}$, Markus Herrmann-Wicklmayr$^{2}$,\\Kathrin Fla\ss kamp$^{2}$, Peter Eberhard$^{1}$, Henrik Ebel$^{3}$
\thanks{$^{1}$David Leprich, Mario Rosenfelder, and Peter Eberhard are with the Institute of Engineering and Computational Mechanics, University of Stuttgart, 70569 Stuttgart, Germany
        {\tt\small peter.eberhard@itm.uni-stuttgart.de}}%
\thanks{$^{2}$Markus Herrmann-Wicklmayr and Kathrin Fla\ss kamp are with the Chair of Systems Modeling and Simulation, Saarland University, 66123 Saarbrücken, Germany
        {\tt\small kathrin.flasskamp@uni-saarland.de}}%
\thanks{$^{3}$Henrik Ebel is with the Department of Mechanical Engineering, LUT University, Yliopistonkatu 34, 53850 Lappeenranta, Finland
        {\tt\small henrik.ebel@lut.fi}}%
\thanks{This work was supported by the Deutsche Forschungsgemeinschaft (DFG, German Research Foundation)
        under project 501890093, EB195/40-1 “Mehr Intelligenz wagen - Designassistenten in Mechanik und Dynamik
        (SPP 2353)” and under Germany's Excellence Strategy - EXC 2075 - 390740016, project PN4-4 “Learning
        from Data - Predictive Control in Adaptive Multi-Agent Scenarios”.
        K.F.~acknowledges support by the Deutsche Forschungsgemeinschaft (DFG, German Research Foundation) under projects 501928699 and 564445282.}
}

\begin{document}
\tdplotsetmaincoords{55}{25}

\maketitle
\thispagestyle{empty}
\pagestyle{empty}

\begin{abstract}
This article proposes a modular optimal control framework for local three-dimensional ellipsoidal obstacle avoidance, exemplarily applied to model predictive path-following control. Static as well as moving obstacles are considered.
Central to the approach is a computationally efficient and continuously differentiable condition for detecting collisions with ellipsoidal obstacles. 
A novel two-stage optimization approach mitigates numerical issues arising from the structure of the resulting optimal control problem.
The effectiveness of the approach is demonstrated through simulations and real-world experiments with the \Crazyflie quadrotor. 
This represents the first hardware demonstration of an MPC controller of this kind for UAVs in a three-dimensional task.
\end{abstract}
\section{INTRODUCTION}
In real-world robotic applications, collision avoidance is a crucial necessity for safe and autonomous operations.
While numerous methods build upon artificial potential fields~\cite{Khatib86} and control barrier functions~\cite{AmesCooganEgerstedtEtAl19a}, they typically modify the control inputs of a preexisting controller that does not explicitly account for obstacles~\cite{RosenfelderCariusHerrmannWicklmayrEtAl25}.
Integrating collision avoidance into optimization-based motion planning and control is a non-trivial task.
In particular, path-following methods are a common approach to effectively deal with a wide variety of robotic tasks, e.g., in surveillance, logistics, or agriculture~\cite{HungRegoQuintasEtAl23,RubiPerezMorcego20a} but are prone to collisions in complex environments. 
Defining a task by means of a geometric path is intuitive and straightforward for human operators, since, in comparison to defining time-dependent trajectories, a geometric path is time-independent and, therefore, considering the dynamics of the system is less important.
Often, a path-planning algorithm is used to generate a geometric path for a specific task.
Typically, in robotics practice, a higher-level path-planning algorithm, often operating on a purely geometric level, is responsible for providing such collision-free paths.
However, in complex environments, the path-planning algorithm cannot always account for all relevant obstacles, particularly when there are (often smaller-scale) a-priori unknown obstacles discovered on-the-go, or when obstacles exhibit dynamic behavior. 
Even if the robot is equipped with onboard sensors capable of detecting such obstacles, continuously re-planning a global path in real-time often imposes prohibitive computational demands. 
Moreover, if robots additionally have kinematic constraints (e.g., nonholonomic constraints), it may not be possible to exactly follow a path purely planned based on geometrical considerations, so that a local dynamics-aware refinement of planned motions can be necessary, and should, ideally, be aware of obstacles locally. 
However, it is computationally non-trivial to realize a controller that is fast enough for closed-loop real-time operation while accounting for system dynamics and obstacles. 
Since obstacles can be understood as constraints in a robot's configuration space, optimal control approaches like model predictive control~(MPC) seem like a good candidate for the task as they can take into account both (often nonlinear) models of robot dynamics and constraints such as obstacles. 
However, it is non-trivial to include anti-collision constraints in optimal control problems~(OCPs) in a differentiable and quick-to-evaluate manner, so that na\"ive implementations are typically slow and unreliable. 
This article overcomes this issue by building upon a recently proposed local obstacle-avoiding setpoint-tracking MPC controller~\cite{RosenfelderCariusHerrmannWicklmayrEtAl25}, similar to \cite{GaoMessererDuijkerenEtAl24}. 
Central to the foundational approach is a computationally efficient and continuously differentiable condition for detecting collisions between ellipsoidal obstacles. 

The two main contributions of this article are the extension of
the aforementioned works~\cite{RosenfelderCariusHerrmannWicklmayrEtAl25,GaoMessererDuijkerenEtAl24} to path tracking and to three-dimensional obstacle avoidance in practice, i.e., going beyond the planar case for the first time.
In particular, static as well as moving obstacles are considered. 
Furthermore, numerical issues arising from the structure of the resulting optimal control problem (OCP) are mitigated by a novel two-stage optimization approach, helping to also keep computation times short enough in the three-dimensional case.
Finally, the effectiveness of the approach is demonstrated in real-world experiments, utilizing the \Crazyflie quadrotor. 
To the best of the authors' knowledge, this is the first time that an MPC controller of this kind is demonstrated in real-time in hardware experiments with an unmanned aerial vehicle~(UAV) in a three-dimensional task. 

The article is organized as follows.
First, an efficient test for the overlap of two ellipsoids is recapitulated in~\Cref{sec:CollisionTest}, which serves as an algorithmic primitive for formulating this article's obstacle-avoidance constraints.
Subsequently, the collision-avoidance path-following MPC formulation together with a two-stage optimization approach is stated in \Cref{sec:MPC}. 
Next, the dynamic model of the quadrotor used in the experiments is introduced in \Cref{sec:DynamicModel}.
Finally, the experimental results are presented in \Cref{sec:Results} and a conclusion is drawn in \Cref{sec:Conclusion}.
\section{An Efficient Collision Test for Ellipsoidal Obstacles}\label{sec:CollisionTest}
Detecting collisions between two objects is, in general, a non-trivial task that arises in various fields such as robotics, computer graphics, and physics simulations~\cite{Ericson20}. 
In this work, we restrict our attention to the case of ellipsoidal obstacles in three-dimensional space.
However, the framework could be straightforwardly extended to higher-dimensional hyperellipsoids.
Accordingly, both the robots and the obstacles are represented by ellipsoids $\mathcal{E}(\bm{P},\bm{r}) \subseteq \mathbb{R}^3$, defined through a quadratic form as
\begin{equation}\label{eq:DefinitionEllipsoid}
    \mathcal{E}(\bm{P},\bm{r}) = \{\bm{x} \in \mathbb{R}^3 \,\vert\, (\bm{x}-\bm{r})^\top \bm{P}(\bm{x}-\bm{r}) \leq 1 \},
\end{equation}
where $\bm{P} \in \mathbb{S}_{++}^{3}$ is a symmetric positive-definite matrix that characterizes the ellipsoid's shape, size, and orientation, and $\bm{r} \in \mathbb{R}^3$ specifies its center.
Collision detection between a robot and an obstacle can then be formulated as checking whether the corresponding ellipsoidal sets, which may serve as outer approximations of the actual shapes, intersect.
Note, that requiring $\bm{P} \in \mathbb{S}_{++}^{3}$ might be overly restrictive, as the following approach can also be applied to positive semi-definite matrices $\bm{P} \in \mathbb{S}_{+}^{3}$, allowing obstacles to be defined by degenerate ellipsoids~\cite{RosSabaterThomas02}, such as cylinders.
This is practically relevant, for example, if a UAV is not allowed to fly over certain areas.

\begin{figure}[ht]
\tikzsetnextfilename{KComparison/KComparison}
\centering
\includegraphics{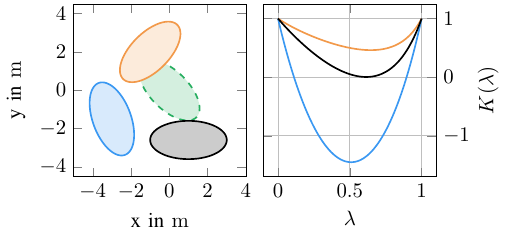}
\caption{Illustration of the function $K(\lambda)$ for different positions of a robot (color coded) relative to the obstacle (green dashed outline).}
\label{fig:KComparison}
\end{figure}

A computationally efficient approach for testing for intersections of ellipsoidal sets, originally introduced in~\cite{RosSabaterThomas02,GilitschenskiHanebeck14} and recently applied in two-dimensional collision avoidance for mobile robots in~\cite{RosenfelderCariusHerrmannWicklmayrEtAl25}, is summarized in the following.  
Given are two ellipsoids $\mathcal{E}_\mathrm{A}(\bm{A},\bm{v})$ and $\mathcal{E}_\mathrm{B}(\bm{B},\bm{w})$ with $\bm{A}, \bm{B} \in \mathbb{S}_{++}^{3}$ and $\bm{v}, \bm{w} \in \mathbb{R}^3$. Testing for the intersection of $\mathcal{E}_\mathrm{A}$ and $\mathcal{E}_\mathrm{B}$ relies on constructing an auxiliary ellipsoid $\mathcal{E}_{\lambda}$ obtained as a convex combination of $\bm{A}$ and $\bm{B}$. 
Specifically, one defines
\begin{subequations}%
\begin{align}
    \hspace{-.5cm} \mathcal{E}_\lambda \coloneqq& \{\bm{x} \in \mathbb{R}^3 \mid (\bm{x}-\bm{m}_\lambda)\tran \bm{E}_\lambda (\bm{x}-\bm{m}_\lambda) \leq K(\lambda) \}, \label{eq:AuxiliaryEllipsoid}\\
    \bm{E}_\lambda &\coloneqq \lambda \bm{A} + (1-\lambda)\bm{B}, \quad \lambda \in [0,1],\\
    \bm{m}_\lambda &\coloneqq \bm{E}_\lambda^{-1}\big(\lambda \bm{A}\bm{v} + (1-\lambda)\bm{B}\bm{w}\big),\\
    \hspace{-.5cm} K(\lambda) &\coloneqq 1 - \lambda \bm{v}\tran \bm{A}\bm{v} - (1-\lambda)\bm{w}\tran\bm{B}\bm{w}
        + \bm{m}_\lambda\tran \bm{E}_\lambda \bm{m}_\lambda, \label{eq:K}
\end{align}
\end{subequations}
where $\mathcal{E}_\lambda$ always satisfies the containment property
\begin{equation}
    (\mathcal{E}_\mathrm{A} \cap \mathcal{E}_\mathrm{B}) \subseteq \mathcal{E}_\lambda \subseteq (\mathcal{E}_\mathrm{A} \cup \mathcal{E}_\mathrm{B}).
\end{equation}
The auxiliary ellipsoid $\mathcal{E}_\lambda$ is always contained within the union of $\mathcal{E}_\mathrm{A}$ and $\mathcal{E}_\mathrm{B}$, whereas the intersection of $\mathcal{E}_\mathrm{A}$ and $\mathcal{E}_\mathrm{B}$ is always contained within $\mathcal{E}_\lambda$, i.e., for every $\lambda\in[0, 1]$, see the proofs in~\cite{GilitschenskiHanebeck14}.
Note that, whereas the union of $\mathcal{E}_\mathrm{A}$ and $\mathcal{E}_\mathrm{B}$ is never empty, the intersection is, by definition, empty if and only if no collision occurs.
Therefore, one possibility to enforce collision avoidance, i.e., $\mathcal{E}_\mathrm{A} \cap \mathcal{E}_\mathrm{B} = \emptyset$, is to ensure that it holds that $\mathcal{E}_\lambda = \emptyset$ for an arbitrary $\lambda \in [0,1]$.
The set $\mathcal{E}_\lambda$ is empty if and only if $K(\lambda)$ is negative for the specified $\lambda$ as $\bm{E}_\lambda$ is always positive-definite as the sum of two positive-definite matrices, leading to an empty set in the definition of the ellipsoid in~\eqref{eq:AuxiliaryEllipsoid} if the right-hand side of the inequality therein becomes negative. 
To conclude, the non-intersection of two ellipsoids can be tested by the condition
\begin{equation}\label{eq:CollisionTest}
    \mathcal{E}_\mathrm{A} \cap \mathcal{E}_\mathrm{B} = \emptyset 
    \iff \exists \lambda \in [0,1] : K(\lambda) < 0.
\end{equation}
The collision test is illustrated in \Cref{fig:KComparison} using a two-dimensional example.
The right panel shows different shapes of $K(\lambda)$ for various positions of a robot relative to a static obstacle. 
The robot's shape and positions are represented by ellipsoids colored blue, orange, and purple, whereas the static obstacle is shown as a green dashed outline in the left panel.
For the blue ellipsoid, no collision occurs, as indicated by $K(\lambda)$ attaining negative values. 
In contrast, the orange ellipsoid corresponds to a collision with $K(\lambda)$ remaining strictly positive for all $\lambda \in [0,1]$. 
A special case is illustrated by the purple ellipsoid, where the two sets are touching, but not overlapping.
Here, $K(\lambda)$ is non-positive at exactly one point $\lambda\in[0,1]$. 

Whereas this article directly employs $K(\lambda)$ from \eqref{eq:CollisionTest} for collision-checking in OCPs, there is also an interesting relationship to the Minkowski sum of ellipsoids, which can also be used to formulate collision-avoidance constraints, as described subsequently. 
The function $K(\lambda)$ given in~(\ref{eq:K})
can equivalently be expressed as
\begin{equation}\label{eq:KLambdaEquivalent}
K(\lambda) = 1 - (\bm{w} - \bm{v})\tran\left(\frac{\bm{B}^{-1}}{1-\lambda} + \frac{\bm{A}^{-1}}{\lambda}\right)^{-1}(\bm{w} - \bm{v})
\end{equation}
as stated in \cite{GilitschenskiHanebeck14}, providing a conceptual link to~\cite{GaoMessererDuijkerenEtAl24}. 
Concretely, whereas the derivation of $K(\lambda)$ originates from the propagation and fusion of two ellipsoids, see~\cite{RosSabaterThomas02}, the equivalent formulation in~\eqref{eq:KLambdaEquivalent} highlights its relation to the Minkowski sum of $\mathcal{E}_\mathrm{A}$ and $\mathcal{E}_\mathrm{B}$.
In theory, the Minkowski sum, defined as
\begin{equation}
    \mathcal{E}_\mathrm{A} \oplus \mathcal{E}_\mathrm{B} \coloneqq \{\bm{x} + \bm{y} \ \vert\ \bm{x} \in \mathcal{E}_\mathrm{A}, \bm{y} \in \mathcal{E}_\mathrm{B}\},
\end{equation}
can be used to test whether $\mathcal{E}_\mathrm{A}$ and $\mathcal{E}_\mathrm{B}$ are intersecting.
This is achieved by checking if the displacement vector $\bm{\eta} \coloneqq \bm{w}-\bm{v}$ between the centers of $\mathcal{E}_\mathrm{A}$ and $\mathcal{E}_\mathrm{B}$ lies within their Minkowski sum, i.e., $\bm{\eta} \in (\mathcal{E}_\mathrm{A} \oplus \mathcal{E}_\mathrm{B})$.
Unfortunately, computing the Minkowski sum of two ellipsoids is nontrivial in practice and does not generally yield another ellipsoid~\cite{YanChirikjian15}.
To mitigate this issue, the ellipsoid $\mathcal{E}\left(\left(\nicefrac{\bm{B}^{-1}}{1-\lambda} + \nicefrac{\bm{A}^{-1}}{\lambda}\right)^{-1}, \bm{w}\right)$ can be used as an over-approximation of the Minkowski sum $\mathcal{E}_\mathrm{A} \oplus \mathcal{E}_\mathrm{B}$ for $\lambda \in [0,1]$, see~\cite{DurieuWalterPolyak01}.
The displacement vector $\bm{\eta}$ not being contained in the over-approximation of the Minkowski sum then implies that $\mathcal{E}_\mathrm{A}$ and $\mathcal{E}_\mathrm{B}$ are not intersecting, which is equivalent to requiring $K(\lambda) < 0$ for some $\lambda \in [0,1]$ with $K(\lambda)$ from~\eqref{eq:KLambdaEquivalent}.
So far, this collision test, alternative to the one used in this paper, utilizes the over-approximation of the Minkowski sum, therefore providing a possibly too conservative result.
However, this over-approximation can be made tight in an arbitrary direction based on $\lambda$~\cite{Houska11,GaoMessererDuijkerenEtAl24}. 
This alternative formulation based on the over-approximation of the Minkowski sum has been applied in \cite{GaoMessererDuijkerenEtAl24}, independently of \cite{RosenfelderCariusHerrmannWicklmayrEtAl25}, to address the ellipsoidal collision avoidance problem in two dimensions for a wheeled mobile robot.
\section{COLLISION-AVOIDANT MODEL PREDICTIVE PATH-FOLLOWING CONTROL}\label{sec:MPC}
We consider the scenario in which the plant shall follow a geometric path using a model predictive controller.  
The task is to navigate through a complex environment containing two types of obstacles, a priori known \emph{global} obstacles of arbitrary shape, and a priori unknown \emph{local} obstacles of ellipsoidal shape.
The global path is computed a priori in the output space of the plant, which has been successfully applied for different type of plants, e.g. industrial robots~\cite{FaulwasserWeberZometaEtAl17}, mobile robots~\cite{FaulwasserFindeisen16}, or quadrotors~\cite{LeprichRosenfelderHermleEtAl25}.

\subsection{PATH-FOLLOWING MPC FORMULATION}
A global path planner is utilized to generate a geometric path~$\mathcal{P}$, defined as
\begin{equation}
    \mathcal{P} \coloneqq \{\bm{p}(s) \in \mathcal{Y} \subseteq \mathbb{R}^{n_\mathrm{p}} \mid s \in [s_0, 0]\},
\end{equation}
where~$s_0 \leq 0$ denotes the path's start, and $n_\mathrm{p} \in \mathbb{N}$ the output dimension, respectively.
The planner accounts for global obstacles by operating exclusively within the feasible output space~$\mathcal{Y}$ of the plant.  
Hence, the path is described by a parametric function~$\bm{p}(s)$ with~$s$ denoting the path parameter.

To convert the time-independent path~$\mathcal{P}$ into a time-dependent trajectory, a timing law~$\bm{g}(\bm{z},\nu)$ is introduced to govern the evolution of the path parameter~$s$.  
In this work, the timing law is chosen as an integrator chain of length two, given by
\begin{equation}\label{eq:TimingLaw}
    \underbrace{\begin{bmatrix}
        \dot{s} \\ \ddot{s}
    \end{bmatrix}}_{\displaystyle\dot{\bm{z}}} = 
    \begin{bmatrix}
        0 & 1 \\
        0 & 0 
    \end{bmatrix}
    \underbrace{\begin{bmatrix}
        s \\ \dot{s}
    \end{bmatrix}}_{\displaystyle\bm{z}} +
    \begin{bmatrix}
        0 \\ 1
    \end{bmatrix}\nu \eqqcolon \bm{g}(\bm{z},\nu),
\end{equation}
where the virtual input~$\nu$ enables the controller to actively regulate the time evolution of the path parameter. 
The latter serves as the resulting reference point on the path.
To now track the geometric path~$\mathcal{P}$ while avoiding local obstacles,
the discrete-time path-following MPC formulation from \cite{FaulwasserFindeisen16} is adapted to derive
\begin{minipage}{\linewidth}
\begin{subequations}\label{eq:MPPFC}%
\begin{align}
    \underset{\bm{u}_{\cdot\mid t},\,\nu_{\cdot\mid t},\,\lambda_{\cdot\mid t}}{\mathrm{min}} \quad 
    & \sum_{k=0}^{N-1} \ell\!\left(\bm{y}_{k\mid t},\,\bm{z}_{k\mid t},\,\bm{u}_{k\mid t},\,\nu_{k\mid t}\right) 
\end{align}
\begin{align}
    \mathrm{subject~to} \quad 
    \bm{x}_{k+1\mid t} &= \bm{f}^{\text{d}}\!\left(\bm{x}_{k\mid t},\,\bm{u}_{k\mid t}\right), && k\in\mathbb{I}_{0:N-1} \label{eq:DynamicConstraint}\\
    \bm{x}_{k\mid t} &\in \mathcal{X}, && k\in\mathbb{I}_{0:N} \label{eq:StateConstraint}\\ 
    \bm{u}_{k\mid t} &\in \mathcal{U}, && k\in\mathbb{I}_{0:N-1} \label{eq:InputConstraint}\\
    \bm{y}_{k\mid t} &= \bm{h}(\bm{x}_{k\mid t}) \in \mathcal{Y}, && k\in\mathbb{I}_{0:N}\\
    \bm{x}_{0\mid t} &= \bm{x}(t), \label{eq:InitialConditionX}\\[0.5\baselineskip]
    \bm{z}_{k+1\mid t} &= \bm{g}^{\text{d}}\!\left(\bm{z}_{k\mid t},\,\nu_{k\mid t}\right), && k\in\mathbb{I}_{0:N-1} \label{eq:TimingLawConstraint}\\
    \bm{z}_{k\mid t} &\in \mathcal{Z}, && k\in\mathbb{I}_{0:N} \label{eq:PathParameterConstraint}\\
    \nu_{k\mid t} &\in \mathcal{V}, && k\in\mathbb{I}_{0:N-1} \label{eq:VirtualInputConstraint}\\
    \bm{z}_{0\mid t} &= \bm{z}(t), \label{eq:InitialConditionZ}\\[0.5\baselineskip]
    K\!\left(\lambda_{k\mid t},\,x_{k\mid t}\right) &\leq 0, && k\in\mathbb{I}_{0:N} \label{eq:CollisionConstraint} \\
    \lambda_{k\mid t} &\in [0,1], && k\in\mathbb{I}_{0:N} \label{eq:LambdaConstraint} 
\end{align}
\end{subequations}
\end{minipage}
with
\begin{align} \label{eq:StageCost}
    \ell\!\left(\bm{y}_{k\mid t},\,\bm{z}_{k\mid t},\,\bm{u}_{k\mid t},\,\nu_{k\mid t}\right) &= \left\lVert\begin{bmatrix}\bm{y}_{k\mid t} - \bm{p}(s_{k\mid t}) \\ s_{k\mid t} \\\bm{u}_{k\mid t} \\ \nu_{k\mid t}\end{bmatrix}\right\rVert_{\bm{W}}^{2} ,
\end{align}
where $\lVert\bm{x}\rVert^2_{\bm{W}}=\bm{x}\tran\bm{Wx}$ and $\mathbb{I}_{a:b} = \{a, a+1, \ldots, b\}$ for~$a,b\in\mathbb{N}_0$ with~$a<b$.
Further, the approximated discrete-time dynamics of the plant and the timing law~(\ref{eq:TimingLaw}), each discretized with a sample time~$\delta$, are denoted by~$\bm{f}^{\text{d}}$ and~$\bm{g}^{\text{d}}$.
The state and inputs are constrained by the given sets~$\bm{x}\in\mathcal{X} \subseteq \mathbb{R}^{n_\mathrm{x}}$ and~$\bm{u}\in\mathcal{U} \subseteq \mathbb{R}^{n_\mathrm{u}}$, with~$n_\mathrm{x}\in\mathbb{N}$ and~$n_\mathrm{u}\in\mathbb{N}$ describing the state and input dimensions of the plant.
The dynamics of the path parameter is constrained to~$\mathcal{Z} = [s_0, 0] \times (0, \dot{s}_\mathrm{max}]$ with~$\dot{s}_\mathrm{max} > 0$ ensuring that the path parameter progresses forward along~$\mathcal{P}$.
The virtual input is constrained by box constraints~$\mathcal{V} = [\nu_\mathrm{min}, \nu_\mathrm{max}]$ with~$\nu_\mathrm{min} < 0 < \nu_\mathrm{max}$. 
Furthermore, the weighting matrix is structured as~$\bm{W}=\operatorname{blockdiag}(\bm{W}_y,\bm{W}_s,\bm{W}_u,\bm{W}_\nu)$ with the output weighting matrix~$\bm{W}_y \in \mathbb{S}_{++}^{n_\mathrm{p}}$, the path parameter weighting matrix~$\bm{W}_s \in \mathbb{S}_{++}$, the control weighting matrix~$\bm{W}_u \in \mathbb{S}_{++}^{n_\mathrm{u}}$ and the virtual input weighting matrix~$\bm{W}_\nu  \in \mathbb{S}_{++}$.
The notation $\bm{x}_{k\mid t}$ in~(\ref{eq:MPPFC}) emphasizes two time concepts, the time step $k$ in the prediction horizon $\mathbb{I}_{0:N}$, and the experiment time $t$. Therefore, $\bm{x}_{k\mid t}$ describes the state prediction at time $t + k\delta$, based on measurements at time $t$.

\subsection{PARAMETERIZED COLLISION-AVOIDANT OCP}
To avoid local obstacles, the collision-avoidance test condition from~\eqref{eq:CollisionTest} is incorporated into the MPC formulation via constraints~(\ref{eq:CollisionConstraint}) and~(\ref{eq:LambdaConstraint}).
Compared to~\eqref{eq:CollisionTest}, these constraints are relaxed, allowing for cases where~$K(\lambda) = 0$ due to floating-point arithmetic.
This corresponds to the special case in which the two ellipsoids are touching, but not overlapping, as illustrated in~\Cref{fig:KComparison}.
Methods to enforce strict negativity of constraint~(\ref{eq:CollisionConstraint}) are discussed in~\Cref{sec:Results}.
Furthermore, the notation in constraint~(\ref{eq:CollisionConstraint}) emphasizes that~$K$ depends on both the state~$\bm{x}_{k\mid t}$ and the decision variable~$\lambda_{k\mid t}$ across the horizon, facilitating predictive motion planning for collision avoidance. 
Integrating multiple local obstacles is straightforward, as each obstacle introduces an additional collision-avoidance parameter $\lambda_l$ and a set of constraints~(\ref{eq:CollisionConstraint}) and~(\ref{eq:LambdaConstraint}), where~$l\in\mathbb{N}$ denotes the obstacle index.
In the following,  without loss of generality, we consider only a single local obstacle.

Solving the MPC problem formulated in~\eqref{eq:MPPFC} is associated with significant numerical challenges, as observed in both simulations and experiments.  
A key source of these difficulties is the absence of~$\lambda_{k\mid t}$ in the stage cost~(\ref{eq:StageCost}).  
For SQP-type solvers, such structural properties can lead to ill-conditioning of the resulting nonlinear program and, consequently, to feasibility issues.  
In~\cite{GaoMessererDuijkerenEtAl24}, the same phenomenon appears for an MPC controller applied to collision-avoidant trajectory tracking of a wheeled mobile robot, and the appearing numerical issues are addressed by regularizing the affected Hessian blocks.  
While this approach resolves numerical difficulties, the regularized problem can still exhibit high computation times, making it unsuitable for real-time applications~\cite{GaoMessererDuijkerenEtAl24}.

A key insight is that solving the OCP~(\ref{eq:MPPFC}) with the variables~$\lambda_{k\mid t}$ predefined over the prediction horizon, i.e., treating at time~$t$ each~$\lambda_{k\mid t}$, $k\in\mathbb{I}_{0:N}$, as a parameter~$\bar{\lambda}_{k\mid t}$ rather than a decision variable, significantly enhances practically observed numerical stability.
The resulting parameterized OCP 
is given by
\begin{equation}
    \label{eq:OCP_fixed_lambda}
    \begin{aligned}
        \hspace{-1pt}\underset{\bm{u}_{\cdot\mid t},\,\nu_{\cdot\mid t}}{\mathrm{min}}\!  J(\bm{x}(t), &\bm{z}(t), \bar{\bm{\lambda}}) \coloneqq 
        \sum_{k=0}^{N-1}\! \ell\!\left(\bm{y}_{k\mid t},\,\bm{z}_{k\mid t},\,\bm{u}_{k\mid t},\,\nu_{k\mid t}\right) 
        \\
        \text{subject to} \quad
        &\text{\eqref{eq:DynamicConstraint} to~\eqref{eq:InitialConditionZ}}\mathrm{~and}
        \\
        &K(\bar{\lambda}_{k\mid t}, \bm{x}_{k\mid t}) \leq 0, \quad k\in\mathbb{I}_{0:N},
    \end{aligned}
\end{equation}
where 
$\bar{\bm{\lambda}} = 
\begin{bmatrix}
    \bar{\lambda}_{0\mid t} & \bar{\lambda}_{1\mid t} & \cdots & \bar{\lambda}_{N\mid t}
\end{bmatrix}\tran$ 
is the stacked parameter vector.
Furthermore, we denote as $J^\star(\bm{x}, \bm{z}, \bar{\bm{\lambda}})$ the resulting optimal cost, depending on the initial conditions for $\bm{x}(t)$ and $\bm{z}(t)$ as well as the parameter $\bar{\bm{\lambda}}$, which are assumed to be such that the OCP is initially feasible.

\begin{remark}\label{rem:ConstantLambda}
As a special case, the collision avoidance parameter can be set to a constant value, meaning that it is set constant across the prediction horizon, but varies over time~$t\in\mathbb{R}_{\geq 0}$, i.e., $\bar{\lambda}(t) = \bar{\lambda}_{k\mid t}\;\forall k \in\mathbb{I}_{0:N}$. Alternatively, it can be held constant across the prediction horizon and over time, i.e., $\bar{\lambda} = \bar{\lambda}_{k\mid t}$, for all $ k \in\mathbb{I}_{0:N}$ and all $t \in \mathbb{R}_{\geq 0}$.
Although this guarantees collision avoidance, it possibly results in a conservative behavior regarding the path-following performance, as also illustrated in~\Cref{sec:DynamicModel}. 
\end{remark}

\subsection{TWO-STAGE OPTIMIZATION APPROACH}\label{sub:TwoStageOptimization}
We propose to choose the parameter vector~$\bar{\bm{\lambda}}$ in the OCP~\eqref{eq:OCP_fixed_lambda} as the solution of
\begin{equation}\label{eq:OptimalLambda}
    \underset{\bar{\lambda}_{k\mid t}\in[0,1]}{\text{argmin}}
    K(\bar{\lambda}_{k\mid t},\bar{\bm{x}}_{k\mid t})
\end{equation}
for each $k\in\mathbb{I}_{0:N}$, where~$\bar{\bm{x}}_{k\mid t}$ is a prediction or candidate for the state along the prediction horizon.
Therefore, providing the minimizer of~$K$ gives the MPC the largest room for shaping~$K$ via the state.
Even though~\eqref{eq:OptimalLambda} is formally an optimization problem, it is cheap to compute since~$K$ is convex w.r.t.~$\lambda$~\cite{GilitschenskiHanebeck14}, scalar, and the minimum is searched on a compact set.
Note that with this approach, $\bar{\lambda}_{k\mid t}$ generally does not only vary over time but also within the prediction horizon.

To compute a solution to~\eqref{eq:OptimalLambda}, the current state and its predicted evolution along the horizon must be known. 
The initial state $\bm{x}_{0\mid t}$ is set to the current measurement $\bm{x}(t)$, while for $k \in \mathbb{I}_{1:N-1}$, the predicted state from the MPC's previous time step, i.e., $\bm{x}_{k\mid t} = \bm{x}_{k+1\mid t-\delta}$, serves as a hot-start candidate.
Although $\bar{\lambda}_{k\mid t}$ minimizes $K(\lambda, \bm{x}_{k\mid t})$ for a given state $\bm{x}_{k\mid t}$, it is not guaranteed to remain optimal once the state trajectory is updated through solving 
OCP~\eqref{eq:OCP_fixed_lambda}, as this possibly yields an update on the predicted optimal state trajectory.
Consequently, the corresponding optimizer~$\bar{\lambda}_{k\mid t}$ may change, motivating an iterative procedure. 
The resulting scheme alternates between optimizing~$\bar{\bm{\lambda}}$ and solving the MPC problem and is summarized in~\Cref{alg:TwoStageOptimization}.

\begin{algorithm}[ht]
	\caption{Two-stage optimization in each time step}\label{alg:TwoStageOptimization} 
	\begin{algorithmic}
		\Require~$\lambda^0_{k\mid t}\in\mathbb{R}$, $\bm{x}^0_{k\mid t}\in\mathbb{R}^{n}$, $\forall k\in\mathbb{I}_{0:N}$ 
		\State$i \gets 0$ 
		\While{
			$i < i_\mathrm{max}$ and $T_\mathrm{comp} < \delta$
		} 
			\For{$k = 0$ to~$N$} 
                \State$\lambda^{i+1}_{k\mid t} = \underset{\lambda\in[0,1]}{\text{argmin}}~K(\lambda,\bm{x}_{k\mid t}^{i})$ 
			\EndFor 
			\State$\bm{u}_{\cdot\mid t}^{i+1},\bm{x}_{\cdot\mid t}^{i+1} \gets \verb|solveMPC|(\bm{x}^{i}_{0\mid t},\bar{\lambda}_{\cdot\mid t} = \lambda^{i+1}_{\cdot\mid t})$ 
			\If{$\forall k\in\mathbb{I}_{0:N}:\vert\lambda^{i+1}_{k\mid t} - \lambda^i_{k\mid t}\vert < \varepsilon$}
            \State \textbf{break}
            \EndIf
            \State$i \gets i + 1$ 
		\EndWhile 
		\State$\bm{u}_\mathrm{mpc} \gets \bm{u}_{0\mid t}^{i}$
	\end{algorithmic} 
\end{algorithm}
Similar to a standard MPC loop, the two-stage algorithm is executed at every control step. 
Initially, guesses for $\lambda_{k\mid t}$ and $\bm{x}_{k\mid t}$ are provided, typically reusing solutions from the previous time step, and the iteration counter is initialized with $i = 0$. 
In each iteration, the optimization~(\ref{eq:OptimalLambda}) is solved for all $k \in \mathbb{I}_{0:N}$, yielding updated parameters $\lambda^{i+1}_{k\mid t}$. 
Subsequently, the 
OCP~\eqref{eq:OCP_fixed_lambda}
is solved using these updated parameters and the current initial state $\bm{x}^{i}_{0\mid t}$. 
This OCP can be solved by any suitable numerical solver, abstracted here as \verb|solveMPC()|. 
Note that one might have additional variables depending on the MPC scheme, such as parameters or variables from added artificial dynamics, as in the path-following scheme introduced in~\eqref{eq:MPPFC}, which are omitted in~\Cref{alg:TwoStageOptimization} without loss of generality.
The resulting state trajectory $\bm{x}^{i+1}_{k\mid t}$ is then used to update $\lambda_{k\mid t}$ in the next iteration. 
The algorithm terminates once the updates in the parameter vector~$\bm{\bar{\lambda}}$ fall below a threshold $\varepsilon$, the maximum number of iterations $i_\mathrm{max}$ is reached, or the computation time exceeds the sampling interval $\delta$. 
Finally, the control input applied to the system is set to $\bm{u}_\mathrm{mpc} = \bm{u}_{0\mid t}^{i}$.
\begin{remark}
	Even if the two-stage optimization in~\Cref{alg:TwoStageOptimization} is terminated before convergence, i.e., the threshold value $\varepsilon$ is not met, the resulting control input still guarantees nominal collision avoidance for static obstacles if the warm start at $i=0$ is feasible.
\end{remark}

Notably, both simulations and real-world experiments indicate that the algorithm in typical real-world scenarios converges sufficiently within a single iteration ($i_\mathrm{max} = 1$). 
While a detailed theoretical convergence analysis remains an open question, the focus in the following is placed on experimental results for the three-dimensional collision-avoidance problem.

\section{Dynamic Model of the Crazyflie}\label{sec:DynamicModel}
\begin{figure}[th]
    \centering
    \includegraphics[scale=0.048]{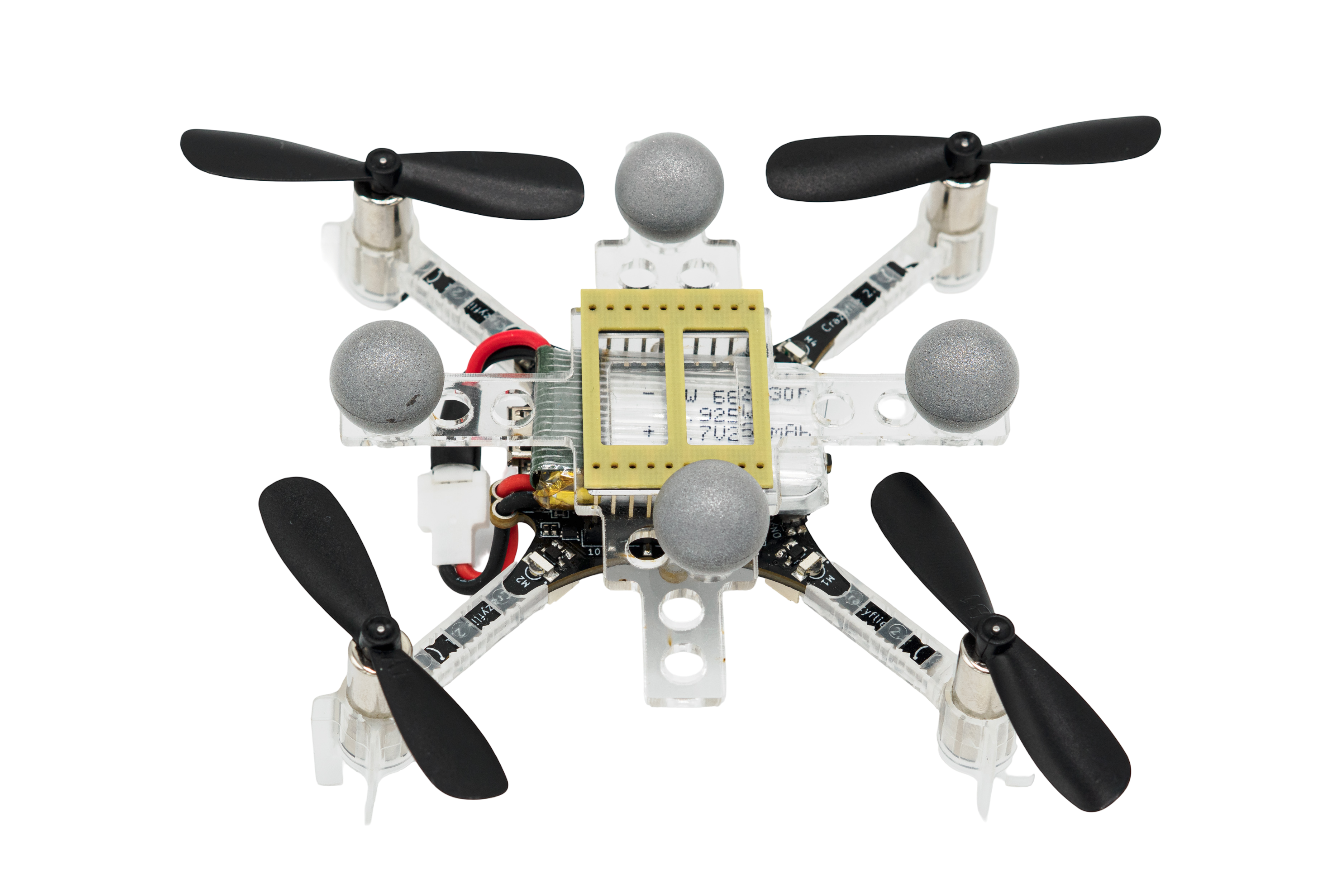}
    \caption{\Crazyflie quadrotor with OptiTrack markers for position and attitude estimation during real-world experiments.}
    \label{fig:Crazyflie}
\end{figure}

Throughout this work, the \Crazyflie quadrotor, shown in \Cref{fig:Crazyflie}, is used to demonstrate the proposed control approach in real-world experiments.
Due to its small size and weight, the \CrazyflieShort is well suited for indoor experiments in confined spaces as demonstrated in \cite{LeprichRosenfelderHermleEtAl25}.  
At the lowest control level, motor commands must be computed at frequencies of up to~$\SI{500}{\hertz}$ to ensure stability, which poses a significant challenge even for modern MPC frameworks.  
To address this, the platform's onboard attitude controllers are employed as fast low-level controllers.  
In contrast, the proposed collision-avoidant MPC scheme serves as a high-level controller, providing reference signals to the attitude control loop at a frequency of~$\SI{50}{\hertz}$.
Utilizing this so-called separated guidance and control approach, see also~\cite[Figs.~1~\&~3]{LeprichRosenfelderHermleEtAl25}, the dynamics imposed by the attitude control loop on the \CrazyflieShort needs to be considered in the high-level MPC formulation.
This work builds upon the continuous-time model
\begin{equation}\label{eq:Dynamics}
	\dot{\bm{x}} = \bm{f}(\bm{x},\bm{u}) = \begin{bmatrix}
        \bm{\dot{\xi}} \\[0.5em]
        \left(\mathrm{s}_\phi \mathrm{s}_\psi + \mathrm{c}_\phi \mathrm{c}_\psi \mathrm{s}_\theta\right)\left(\frac{\Delta T}{m} + g\right)\\[0.5em]
        \left(\mathrm{c}_\phi \mathrm{s}_\psi \mathrm{s}_\theta - \mathrm{c}_\psi \mathrm{s}_\phi\right)\left(\frac{\Delta T}{m} + g\right)\\[0.5em]
        -g + \mathrm{c}_\phi \mathrm{c}_\theta\left(\frac{\Delta T}{m} + g\right)\\[0.5em]
        \frac{1}{T_\phi}(\phi_\mathrm{cmd} - \phi) \\[0.5em]
        \frac{1}{T_\theta}(\theta_\mathrm{cmd} - \theta) \\[0.5em]
        \dot{\psi}_\mathrm{cmd}   
    \end{bmatrix}
\end{equation}
to represent the \CrazyflieShort dynamics for control purposes, where~$\mathrm{s}_x$ and~$\mathrm{c}_x$ denote~$\sin{x}$ and~$\cos{x}$, respectively.  
The position of the drone in the inertial frame is~$\bm{\xi} = \begin{bmatrix}x & y & z\end{bmatrix}\tran$ and the orientation is~$\bm{\Theta} = \begin{bmatrix}\phi & \theta & \psi\end{bmatrix}\tran$, representing roll, pitch, and yaw angles.  
The state vector is defined as\linebreak$\bm{x} = \begin{bmatrix}\bm{\xi}\tran & \dot{\bm{\xi}}\tran & \bm{\Theta}\tran\end{bmatrix}\tran \in \mathcal{X} \subseteq \mathbb{R}^9$, and the input vector as~$\bm{u} = \begin{bmatrix}\Delta T & \phi_\mathrm{cmd} & \theta_\mathrm{cmd} & \dot{\psi}_\mathrm{cmd}\end{bmatrix}\tran \in \mathcal{U} \subseteq \mathbb{R}^4$, where~$\Delta T$ is the deviation of the total thrust~$T$ from the hover thrust~$T_{\text{h}} = m\, g$, i.e.,~$\Delta T = T - T_{\text{h}}$, and the subscript $(\cdot)_{\mathrm{cmd}}$ indicates the setpoints commanded to the attitude controller.  
The system's output is given in terms of~$\bm{y} = \begin{bmatrix}\bm{\xi}\tran & \psi\end{bmatrix}\tran \in \mathcal{Y}$, where~$\mathcal{Y}$ denotes the output space of the drone.  
The influence of the attitude control loop on the rotational dynamics of the \CrazyflieShort is considered by three first-order systems, with time constants~$T_\phi$ and~$T_\theta$ for roll and pitch, respectively, compare~\eqref{eq:Dynamics}.
Furthermore,~$m$ is the mass of the drone and~$g$ the gravitational acceleration.
A model of this kind has been successfully employed in several works, e.g.,  \cite{LeprichRosenfelderHermleEtAl25,LlanesKakishWilliamsEtAl24,HuangBauerPan22,KamelStastnyAlexisEtAl17}.

\begin{figure}[ht]
    \centering
    \tikzsetnextfilename{Path/Path}
    \includegraphics{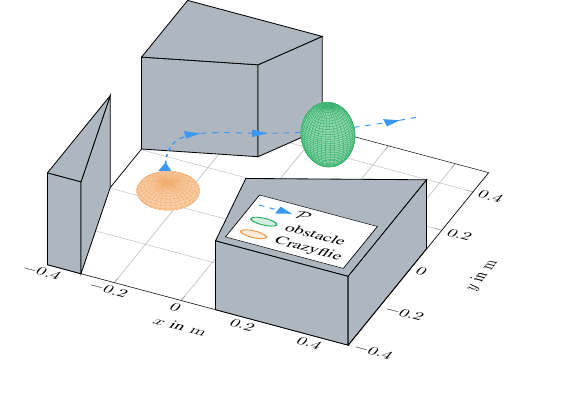}
    \caption{Crazyflie (orange) navigating through a complex environment with a priori known obstacles (grey) and a locally detected, a priori unknown ellipsoidal obstacle (green). The blue dashed lines indicate the reference path~$\mathcal{P}$.}
    \label{fig:Path}
\end{figure}

For the remainder of this paper, the discussion is guided by an exemplary scenario illustrated in~\Cref{fig:Path}.
Assume that the global path planner provides the path
\begin{equation}\label{eq:Path}
    \bm{p}(s) = 
    \begin{bmatrix}
        \tfrac{\sqrt{2}}{2}\Bigl((0.75s + 0.5) - e^{-(6s + 5.8)}(2.25s + 2.175)\Bigr) \\[0.5em]
        \tfrac{\sqrt{2}}{2}\Bigl((0.75s + 0.5) + e^{-(6s + 5.8)}(2.25s + 2.175)\Bigr) \\
        0.5 \\
        \mathrm{arctan}\left(-\frac{4}{30}(135s + 108)e^{-(6s + 5.8)}\right) + \frac{\pi}{4}
    \end{bmatrix}\nonumber
\end{equation}
with~$s_0 = -1$.  
The path~$\mathcal{P}$ is shown in~\Cref{fig:Path}, which is planned to avoid global obstacles (grey). 
A local obstacle (green), unknown to the global planner, intrudes this path and would cause a collision without the use of online collision-avoidance techniques.
The shape of the \CrazyflieShort is approximated by an ellipsoid based on its physical dimensions.  
The drone has a length and width of~$\SI{0.15}{\meter}$ and a height of~$\SI{0.045}{\meter}$.  
Accordingly, the ellipsoid representing the drone is defined by the positive-definite matrix
\begin{equation}\label{eq:shape_drone}
    \bm{A} = \begin{bmatrix}
        177.78 & 0 & 0 \\
        0 & 177.78 & 0 \\
        0 & 0 & 1975.3
    \end{bmatrix}\SI{}{\per\meter\squared} \in \mathbb{S}_{++}^3
\end{equation}
with the center of the ellipsoid $\bm{v}$ located at the drone's center of gravity~$\bm{\xi}$.
It is assumed that the roll and pitch of the drone remains small at all times such that~\eqref{eq:shape_drone} constitutes a valid outer approximation.
Furthermore, the local obstacle is represented by the ellipsoid defined by
\begin{equation}
    \bm{B} = \begin{bmatrix}
        234.57 & -67.42 & 0 \\
        -67.42 & 190.76 & 0 \\
        0 & 0 & 35.44
    \end{bmatrix}\SI{}{\per\meter\squared} \in \mathbb{S}_{++}^3
\end{equation}
with its center located at $\bm{w} = \begin{bmatrix}
    0.2 & 0.16 & 0.5
\end{bmatrix}\tran\SI{}{\meter}$.
The ellipsoids are depicted in~\Cref{fig:Path} in orange and green, respectively. %
Before investigating a real-world experiment using the two-stage optimization scheme proposed in~\Cref{sub:TwoStageOptimization}, we briefly consider the special case of setting all $\bar{\lambda}_{k\mid t}$ to one fixed value, i.e., $\bar{\lambda}_{k\mid t} \eqqcolon \hat{\lambda}$ for all $k\in\mathbb{I}_{0:N}$ and all times~$t\in\mathbb{R}_{\geq0}$, see~\Cref{rem:ConstantLambda}.
In this way, we want to raise awareness to the fact that two different realizations of such fixed~$\bar{\bm{\lambda}}$ can result in significantly different trajectories.

\begin{figure}[ht]
    \centering
    \tikzsetnextfilename{ConservativeLambdas/ConservativeLambdas}
    \includegraphics{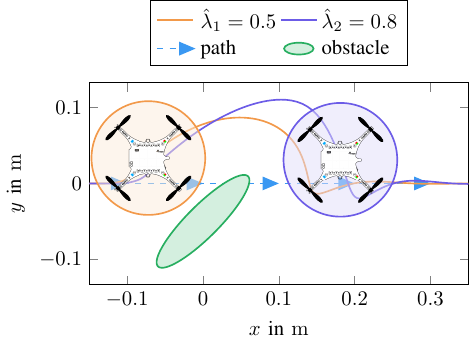}
    \caption{Comparison of two different values~$\hat{\lambda}$ in the collision-avoidance test.}
    \label{fig:ConservativeLambdas}
\end{figure}

As can be seen in \Cref{fig:ConservativeLambdas}, the configuration of path and obstacle forces the drone to deviate from the path in order to avoid a collision with the obstacle.
Nevertheless, ideally, the drone should remain as close as possible to the path, as encoded in the stage cost~(\ref{eq:StageCost}).
Choosing a fixed parameter of $\hat{\lambda}_1 = 0.5$ appears to be less conservative than $\hat{\lambda}_2 = 0.8$.
However, a non-conservative choice of~$\hat{\lambda}$ is not immediately apparent and possibly changes over time, which motivates the proposed two-stage optimization scheme.

\section{EXPERIMENTAL RESULTS}\label{sec:Results}
\begin{figure*}[ht]
    \centering
    \begin{subfigure}[t]{0.47\textwidth}
        \tikzsetnextfilename{Experiments/2DPath}
        \vspace{0.5cm}
        \includegraphics[scale=0.975]{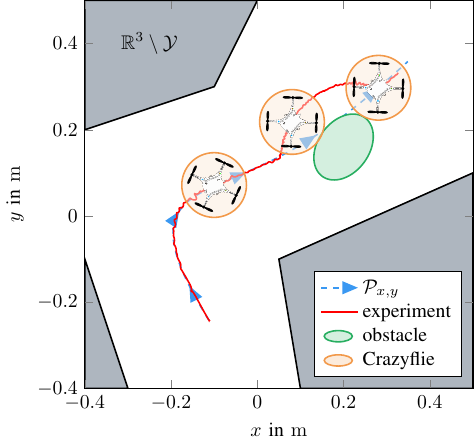}
        \caption{Illustration of the complete experimental result of the \CrazyflieShort for the proposed collision-avoidance MPC applied to the path-following scenario described in~\Cref{fig:Path}.}
        \label{fig:2DPath}
    \end{subfigure}
    \hfill
    \begin{subfigure}[t]{0.47\textwidth}
        \tikzsetnextfilename{Experiments/2DPathExperiment}
        \vspace{0.5cm}
        \includegraphics[scale=0.975]{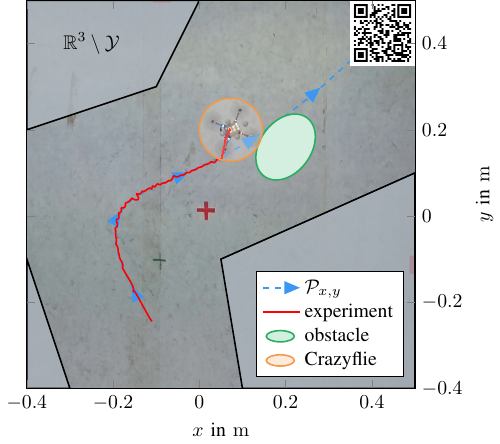}
        \caption{Snapshot of the \CrazyflieShort tracking the path $\mathcal{P}$ (coloured in dashed blue) in a real-world experiment. The past trajectory of the quadrotor, outlined in an orange ellipse, is depicted in red. The local obstacle which is to be avoided is coloured in green.}
        \label{fig:2DPathExperiment}
    \end{subfigure}
    \caption{Results of the real-world experiment. Visualized for different time points in \Cref{fig:2DPath}. A snapshot of the \CrazyflieShort during the experiment is shown in \Cref{fig:2DPathExperiment}.}
    \label{fig:ExperimentalResultPath}
\end{figure*}
In the following section, the performance of the proposed MPC approach is demonstrated in a real-world experiment.
To minimize environmental factors and enable the use of our motion capture systems, the following experiment is conducted in an indoor environment.
The previously introduced \Crazyflie quadrotor is employed as the test platform.
Precise tracking of the quadrotor's position and attitude is achieved by utilizing the \emph{OptiTrack} motion capture system.
Equipped with three \emph{PrimeX 13} and three \emph{Prime 13} cameras in the workspace, the system provides measurements at a frequency of up to $\SI{240}{\hertz}$.
The markers used to track the \CrazyflieShort can be seen in~\Cref{fig:Crazyflie}.
A moving average finite-difference calculation is used to estimate the quadrotor's translational velocity.
In the following, we consider the scenario introduced in~\Cref{sec:MPC}, where a global path-planner provides a geometric path, taking into account a-priori known obstacles.
The \CrazyflieShort is tasked with following this path while avoiding collision with a static ellipsoidal obstacle detected at runtime, as depicted in~\Cref{fig:Path}. 
The results of this experiment are visualized in~\Cref{fig:ExperimentalResultPath}.
In the beginning of each experiment, the quadrotor is maneuvered to the beginning of the path $\bm{p}(s = s_0)$ by utilizing the onboard position controller.
After reaching the start of the path, the MPC controller is activated and takes over control.
The controller is implemented using the \verb|acados| framework~\cite{VerschuerenFrisonKouzoupisEtAl22} and the QP subproblems appearing therein are solved using the \verb|HPIPM|~\cite{FrisonDiehl20} solver.
The prediction horizon is set to $N = 20$ with a time discretization of $\delta = \SI{20}{\milli\second}$.
The two-stage approach, see~\Cref{alg:TwoStageOptimization}, is employed with a maximal iteration count of $i_\mathrm{max} = 1$ to ensure real-time capability with the employed computation hardware.
To solve the convex optimization problem~\eqref{eq:OptimalLambda} of dimension~$N+1$, a simple bisection method for root-finding of its gradient
is employed. 
The bisection is observed to converge in the low microseconds range, requiring only a few iterations with a precision of~$10^{-4}$.

In~\Cref{fig:2DPath}, the \CrazyflieShort is depicted for three different time points during the experiment.
It is observed that the quadrotor successfully tracks the path, colored in dashed blue, while avoiding the local obstacle, colored in green.
The measured trajectory of the \CrazyflieShort is illustrated in red.
The three time points highlighted emphasize the successful collision avoidance, where the orange ellipse, outlining the \CrazyflieShort, and the green ellipse are not overlapping at any time.
This is expected behavior since the OCP~\eqref{eq:OCP_fixed_lambda}
enforces collision-avoidance at all times but simultaneously tries to stay as close as possible to the path.
In~\Cref{fig:2DPathExperiment}, a snapshot of the conducted experiment is illustrated.
In the time point captured, the \CrazyflieShort is currently in contact with the obstacle, making its way around the obstacle while trying to stay as close as possible to the path.
A full video recording of the experiment is available by scanning the QR code in the top right corner of~\Cref{fig:2DPathExperiment}, see also~\cite{LeprichRosenfelderHerrmannWicklmayrEtAl25}.

\begin{figure}[ht]
    \centering
    \tikzsetnextfilename{Experiments/KLambda}
    \includegraphics{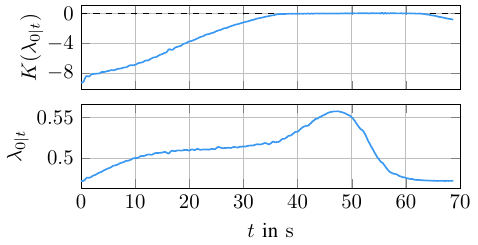}
    \caption{Evolution of $\lambda_{0\mid t}$ and $K(\lambda_{0\mid t})$ during real-world experiment.}
    \label{fig:KLambda}
\end{figure}

To confirm that the \CrazyflieShort and the obstacle are not overlapping during the experiment, the evolution of $K(\lambda_{0\mid t}, \bm{x}(t))$ is illustrated in the upper part of~\Cref{fig:KLambda}.
There, the value of~$K$ becomes zero between $t=\SI{40}{\second}$ and $t=\SI{60}{\second}$, indicating a contact between the two ellipsoids.
However, for a short time, $K$ is slightly positive, which is due to small disturbances and inaccuracies in the state estimation. 
In such scenarios, solver crashes are prevented by implementing constraint~(\ref{eq:CollisionConstraint}) as a soft constraint via the \verb|acados| framework, allowing the solver to trade infeasibility for higher cost function values.
To mitigate this problem, the obstacle can be chosen slightly larger to add a buffer zone.
Alternatively, a small tolerance can be added to the collision constraint~(\ref{eq:CollisionConstraint}), i.e., $K(\bar{\lambda}_{k\mid t}, \bm{x}_{k\mid t}) \leq -\alpha$ with $\alpha > 0$.
The bottom part of~\Cref{fig:KLambda} shows the evolution of $\lambda_{0\mid t}$ over time.
It underlines the benefit of choosing non-constant parameters~$\lambda_{k\mid t}$ w.r.t.\ $t$, as they may vary significantly over time.

\begin{figure}[ht]
    \centering
    \tikzsetnextfilename{Experiments/CostComp}
    \includegraphics{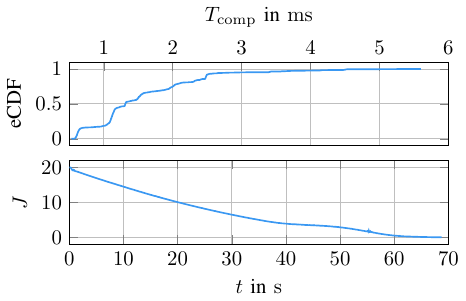}
    \caption{Computation time $T_\mathrm{comp}$ of applying the two-stage optimization algorithm~(\ref{alg:TwoStageOptimization}) as eCDF and the evolution of the cost $J$ over time.}
    \label{fig:TComp}
\end{figure}

The computation time $T_\mathrm{comp}$ of applying the two-stage optimization approach in~\Cref{alg:TwoStageOptimization} is depicted in the upper part of~\Cref{fig:TComp} as an empirical cumulative distribution function~(eCDF).
It is evident that $\SI{75}{\percent}$ of all applications of the two-stage optimization approach are completed within less than $\SI{2}{\milli\second}$.
In total none of the applications exceed the time step~$\delta$ of~$\SI{20}{\milli\second}$ with a maximum computation time of~$\SI{5.6}{\milli\second}$.

\begin{figure}[ht]
    \centering
    \tikzsetnextfilename{DynamicObstacle/DynamicObstacle}
    \includegraphics[scale=0.965]{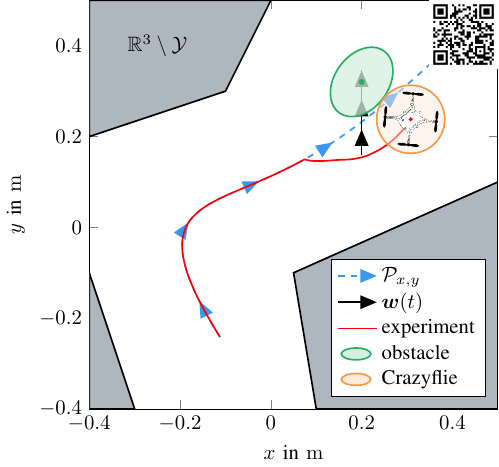}
    \caption{Simulation results of the proposed MPC formulation~(\ref{eq:MPPFC}) applied to a path-following scenario with a dynamic obstacle. The path $\mathcal{P}$ is coloured in dashed blue, the past trajectory of the quadrotor is depicted in red, and the local obstacle is coloured in green. A video of the simulation is available by scanning the QR code in the top right corner.}
    \label{fig:DynamicObstacle}
\end{figure}

The presented MPC formulation~(\ref{eq:MPPFC}) can be adapted to include dynamic obstacles as well, i.e., where the parameters describing the obstacle's ellipsoid are time-variant.
Exemplarily, the same scenario as in the previous experiment is considered. Furthermore, the local ellipsoidal obstacle is now moving with a constant translational velocity of~$\dot{\bm{w}}(t) = \begin{bmatrix}
    0 & 0.005 & 0
\end{bmatrix}\tran\SI{}{\meter\per\second}$.
The simulation results are depicted in~\Cref{fig:DynamicObstacle}, and a video of the simulation is available by scanning the QR code in the top right corner of the figure.  All videos are available in~\cite{LeprichRosenfelderHerrmannWicklmayrEtAl25}.

\section{CONCLUSION}\label{sec:Conclusion}
This paper presented a modular optimal control framework for local three-dimensional obstacle avoidance, exemplarily applied to model predictive path-following control. 
A central contribution is a computationally efficient and continuously differentiable collision detection condition for ellipsoidal obstacles, applicable to both static and moving cases that can be employed in optimal control. 
Numerical challenges arising from the resulting problem structure were addressed through a dedicated two-stage optimization scheme, enabling real-time feasibility. 
The proposed approach was validated in simulation and experimentally on the \Crazyflie quadrotor, demonstrating reliable path tracking and successful avoidance of dynamic and static obstacles. 
To the best of the authors' knowledge, this represents the first real-time hardware implementation of an MPC-based method of this kind for fully three-dimensional UAV scenarios. 
Future research will focus on establishing theoretical convergence guarantees for~\Cref{alg:TwoStageOptimization} and extending the collision detection formulation to encompass more general convex obstacle representations, such as zonotopes.


\end{document}